\documentclass[a4paper,11pt]{article}
\usepackage{jheppub}
\pdfoutput=1
\usepackage{comment}
\usepackage[english]{babel}
\usepackage[utf8x]{inputenc}
\usepackage{amsmath}
\usepackage{amssymb}
\usepackage{graphicx}
\usepackage{subcaption}
\usepackage{caption}
\usepackage{tikz-feynman} 
\newcommand{\pd}[2]{\frac{\partial #1}{\partial #2}}

\newcommand{\be}{\begin{equation}}
\newcommand{\ee}{\end{equation}}
\newcommand{\bea}{\begin{eqnarray}}
\newcommand{\eea}{\end{eqnarray}}
\newcommand{\ba}{\begin{array}}
\newcommand{\ea}{\end{array}}

\begin{document}

%\hfill{ACFI-T21-07}

\title{Tunneling Potentials for the Tunneling Action: Gauge Invariance
%A gauge-independent calculation of the tunneling action using tunneling potentials in the Abelian Higgs theory
}

\preprint{ACFI-T21-07}

\author[a,b]{Suntharan Arunasalam}
\emailAdd{suntharan.arunasalam@gmail.com}
\author[a,b,c,d]{Michael J. Ramsey-Musolf}
\emailAdd{mjrm@sjtu.edu.cn,mjrm@physics.umass.edu}

\affiliation[a]{Tsung-Dao Lee Institute and School of Physics and Astronomy, Shanghai Jiao Tong University, 800 Dongchuan Road, Shanghai, 200240 China}
\affiliation[b]{Shanghai Key Laboratory for Particle Physics and Cosmology, Key Laboratory for Particle Astrophysics and Cosmology (MOE), Shanghai Jiao Tong University, Shanghai 200240, China}
\affiliation[c]{Amherst Center for Fundamental Interactions, Department of Physics, University of Massachusetts, Amherst, MA 01003, USA}
\affiliation[d]{Kellogg Radiation Laboratory, California Institute of Technology, Pasadena, CA 91125 USA}

\abstract
{We formulate a procedure to obtain a gauge-invariant tunneling rate at zero temperature using the recently developed tunneling potential approach. This procedure relies on a consistent power counting in gauge coupling and a derivative expansion. The tunneling potential approach, while numerically more efficient than the standard bounce solution method, inherits the gauge-dependence of the latter when na\"ively implemented. Using the Abelian Higgs model, we show how to obtain a tunneling rate whose residual gauge-dependence arises solely from the polynomial approximations adopted in the tunneling potential computation. }

\maketitle

\section{Introduction}
The topic of phase transitions in the early universe has a long and rich history at the interface of cosmology and particle physics. In recent years, determining the thermal history of symmetry breaking in the Standard Model of particle physics and its possible extensions has seen a resurgence of interest. The extensive experimental program at the Relativistic Heavy Ion Collider has confirmed the results of lattice computations that the transition to the confined phase of quantum chromodynamics at zero baryon chemical potential is a smooth crossover. Analogous lattice computations in the electroweak theory imply that the electroweak symmetry-breaking (EWSB) transition is also a smooth crossover for a Higgs boson heavier than $\sim 70-80$ GeV\cite{Kajantie:1996qd,Laine:1998jb,Csikor:1998eu,Farakos:1994kx,Kajantie:1995kf,Kajantie:1996mn,Farakos:1995dn,Rummukainen:1998as,Gurtler:1997hr,Aoki:1999fi} -- well below the measured value of $125$ GeV. Nevertheless, there exists strong theoretical motivation to consider embedding the Standard Model in a larger beyond the Standard Model (BSM) framework that includes new particles with sub-TeV masses. 

If such an embedding is realized in nature, then it is interesting to ask whether the BSM particles and their interactions would qualitatively change the nature of the EWSB transition and possibly introduce new, earlier phases in the early universe that precede the present Higgs phase. At the same time, the existence of a first order phase transition to the Higgs phase and/or an earlier phase could provide the necessary preconditions for generation of the cosmic baryon asymmetry via electroweak baryogenesis (see Ref.~\cite{Morrissey:2012db} and references therein) while simultaneously giving rise to relic gravitational radiation. The existence of such a thermal history could in principle be inferred indirectly from searches for new particles and/or deviations of Higgs boson properties from Standard Model expectations at the CERN Large Hadron Collider or prospective future colliders\cite{Ramsey-Musolf:2019lsf}. A direct echo of the first order phase transition could detected in next generation gravitational wave searches (see, {\it e.g.}, Refs.~\cite{Caprini:2015zlo,Huang:2016odd,Alves:2018jsw,Caprini:2019egz,Alves:2019igs,Zhou:2020idp}).

In this context, a definitive confrontation of theory and experiment requires the most theoretically robust computations of the early universe thermodynamics in a given BSM scenario and of the corresponding dynamics. The first step is to identify the phase diagram and associated critical temperature $T_C$ for each possible transition from one phase to another. Here, one already encounters a need to go beyond perturbation theory, as the latter cannot ascertain when a given transition is crossover (compared to a {\it bona fide} phase transition). The second step entails analyzing the dynamics. The mere existence of a possible transition does not in itself guarantee that it will have occurred, even if nominally energetically favorable. 

In the case of a first order transition, characterized by the presence of a potential barrier between the two phases, one must also determine that the transition (tunneling) rate is sufficiently large compared to the Hubble rate at the transition temperature. Moreover, determining the frequency spectrum of the associated gravitational radiation requires knowledge of the tunneling rate per unit volume, 
\be
\Gamma=Ae^{-B} 
\label{nucleationdef}
\ee
where at $T=0$ $B$ is the four-dimensional (4d) Euclidean effective action, $S_E$, while for $T>0$ one has $B=S_3/T$, with $S_3$ being the 3d Euclidean action. The prefactor $A$ is generally estimated on dimensional grounds and at finite-$T$ is generally taken to carry a mild $T$-dependence.
As a first order phase transition proceeds via bubble nucleation, one often refers to $\Gamma$ as the nucleation rate, with $T_N$ denoting the temperature at the onset of nucleation, with $T_N < T_C$. 

The classic approach for computing $\Gamma$ at any temperature is via the computation of the bounce action \cite{Coleman:1977py,Callan:1977pt} using the 	\lq\lq overshoot/undershoot\rq\rq\ method. While this approach works particularly well for phase transitions involving a single field, it can be intractable in the context of multi-field problems. Over the past decade, alternate approaches have been developed for this case, including CosmoTransitions\cite{Wainwright:2011kj}, BSMPT\cite{Basler:2018cwe,Basler:2020nrq}, and Bubble Profiler\cite{Athron:2019nbd}. Recently, a powerful algebraic method for computing $B$ in both the single and multi-field contexts was developed in Refs. \cite{Espinosa:2018hue, Espinosa:2018szu}. The method relies on the introduction of an auxiliary function known as the \lq\lq tunneling potential\rq\rq\ , $V_t$, and the use of approximations that facilitate algebraic solution of the field equations to high numerical accuracy without directly solving any differential equations. In principle, the tunneling potential approach holds the promise of facilitating a more efficient survey of nucleation dynamics over a wide array of models and model parameter space as compared to pre-existing methods and numerical packages.

In practice, a common challenge when computing the nucleation rate and $T_N$ is maintenance of gauge invariance. As a physical observable, $\Gamma$ is a gauge-invariant quantity. Formally, this property is encoded in the Nielsen identities\cite{Nielsen:1975fs,Fukuda:1975di}, which state that the effective action is gauge invariant when evaluated using an extremal field configuration. Unfortunately, common computational approaches -- when na\"ively implemented -- do not necessarily respect this requirement. The gauge invariance problem is especially challenging when radiative corrections generate the potential barrier between different vacua. Well-known examples include the Abelian Higgs model as well as the Standard Model. In both cases, it is possible to implement gauge invariance by adopting a well-defined power-counting in the relevant couplings and a systematic truncation procedure\cite{Metaxas:1995ab}. Alternatively, in the presence of a tree-level barrier, it may be desirable to incorporate higher-order corrections -- including gauge boson degrees of freedom --  to the manifestly gauge-invariant tree-level computation. Indeed, recent results for the thermodynamics of the real triplet of the SM, wherein a tree-level barrier can appear, suggests that thermal (loop) corrections can play a more significant role than one might otherwise expect\cite{Niemi:2020hto}. Including these higher order corrections thus introduces the gauge invariance problem. While it is possible to implement these higher order corrections using the $\hbar$-expansion about the tree-level Euclidean action (for relevant discussions, see {\it e.g.}, Refs.~\cite{Laine:1994zq,Patel:2011th}), the practical viability of computing the relevant fluctuation determinant remains to be fully explored. 

In light of the novelty and promise of the tunneling potential approach, we investigate here the implementation of gauge invariance in this framework. {Perhaps, unsurprisingly,  implementation of the tunneling potential method without a well-defined procedure for maintaining gauge invariance will lead to gauge-dependent artifacts. Fortunately, one may circumvent this pitfall by adopting a derivative expansion, as in Ref.~\cite{Metaxas:1995ab,Weinberg:1992ds}. }For concreteness, we adopt the Abelian Higgs model and show how, for $\Gamma(T=0)$, to translate the derivative expansion and power counting of Ref.~\cite{Metaxas:1995ab} into the tunneling potential method. This translation maintains the advantage of the purely algebraic procedure outlined in Refs.~\cite{Espinosa:2018hue, Espinosa:2018szu} and is otherwise relatively straightforward. We expect that this translation will work equally well for any scenario in which one is able to define a leading order, gauge invariant tunneling action -- whether or not the potential barrier exists at tree level or, as in the Abelian Higgs model, occurs through gauge-boson loops.  {Application of these ideas to the tunneling rate at $T>0$ will appear in a future publication}.

{Our discussion is organized as follows}. In Section \ref{model} we present the Abelian Higgs model and the relevant quantities for this approach. {In Section \ref{sec:problem}, we discuss the gauge-dependence problem in the conventional bounce solution context as well as the tunneling potential framework.
Section \ref{derivexpansion} reviews the use of the derivative expansion to obtain a gauge-independent nucleation rate in the bounce solution method.  In Section \ref{tunnellingpotentialapproach}, we translate the derivative expansion into the tunneling potential approach and show how doing so yields a gauge-independent calculation of the tunneling action.  Section \ref{Summary} summarizes our results. Relevant technical details appear in two Appendices. }

\section{The model\label{model}}
%In this paper, we will investigate the gauge dependence of the Abelian Higgs model which consists of a complex scalar that transforms under a $U(1)$ gauge group. 
The Lagrangian for the Abelian Higgs model is 
\begin{align}
    \mathcal{L}=-\frac{1}{4} F_{\mu \nu} F^{\mu \nu}+\frac{1}{2}\left(D_{\mu} \Phi\right)^{*} D^{\mu} \Phi-V_{0}\left(\Phi^{*} \Phi\right)
\end{align}
where $F_{\mu\nu}=\partial_\mu A_\nu-\partial_\nu A_\mu$ and $D_\mu=\partial_\mu-ie A_\mu$ are the standard electromagnetic tensor and covariant derivative respectively,  and where the potential is 
\begin{align}
    V_0(\Phi^*\Phi)=\frac{1}{2}m^2 \Phi^*\Phi+\frac{\lambda}{4}\left(\Phi^*\Phi\right)^2\ \ \ .
\end{align}
We use the background field approach wherein we write the complex scalar as $\Phi=\phi+\frac{1}{\sqrt{2}}(h+i\chi)$ where $\phi$ is the real space-time independent background field and $h$ and $\chi$ are the Higgs and Goldstone bosons, respectively. We now introduce the $R_\xi$-gauge through the gauge-fixing and ghost Lagrangians, 
{
\begin{align}
%    \mathcal{L}_{GF}&=-\frac{1}{2\xi}\left(\partial_\mu A^\mu-\sqrt{2} e\xi\phi\chi\right)^2\\
\mathcal{L}_{GF}&=-\frac{1}{2\xi}\left[\partial_\mu A^\mu-ie\xi\left(\phi^\ast\Phi-\Phi^\ast\phi\right)\right]^2\\
%    \mathcal{L}_{FP}&=\partial_\mu\bar{c}\partial^\mu c -\sqrt{2}e^2\xi h\phi \bar{c}c - 2 e^2\xi \phi^2\bar{c}c
\mathcal{L}_{FP}&=\partial_\mu\bar{c}\partial^\mu c -e^2\xi\left(\phi^\ast\Phi+\Phi^\ast\phi\right)  \bar{c}c \ \ \ .
\end{align}
}
The tree-level masses of these fields and the corresponding numbers of degrees of freedom can be seen in Table \ref{treelevelmass}.

\begin{table}[htbp]
\begin{tabular}{r|c|c}
Field             & Tree level $(\mathrm{mass})^2$ & \#dof  \\
\hline  
Space-like gauge polarization & $2 e^2\phi^2$ & 3   \\
Time-like gauge polarization    & $2\xi e^2\phi^2$ & 1 \\
Higgs boson     & $\frac{1}{2}m^2+\frac{3}{2}\lambda \phi^2$ & 1 \\
Goldstone  boson     & $\frac{1}{2}m^2+\frac{1}{2}\lambda \phi^2+2\xi e^2\phi^2$ & 1\\
Ghost&  $2\xi e^2\phi^2$&-2 \\
\end{tabular}
\caption{The tree level masses and the number of degrees of freedom of the fields}
\label{treelevelmass}
\end{table}

\noindent We now impose a power counting of the form $\lambda\sim e^4$ {and $m^2\sim e^4\,\langle \phi\rangle^2$ as in \cite{Metaxas:1995ab}\footnote{{We believe the statement of power counting for $m^2$ in \cite{Metaxas:1995ab} contains a typo and should read as given here.}}}. Doing so introduces a barrier in the leading order potential, $V_{e^4}$, through loop effects. The  effective potential can  be calculated in the usual way as
\begin{align}
    V_\text{eff}=\sum_\mathrm{j\in particles} -\frac{in_j}{2}\int \frac{d^4k}{(2\pi)^4} \ln(k^2-m_j^2)\ \ \ ,
\end{align}
where $n_j$ are the numbers of degrees of freedom as in Table \ref{treelevelmass} and $m_j$ are the tree level masses. We evaluate these integrals  using the $\overline{\mathrm{MS}}$ renormalization scheme as
\begin{align}
    -\frac{i}{2}\int \frac{d^4k}{(2\pi)^4} \ln(k^2-m_j^2)= \frac{m_j^4}{64\pi^2}\left[\ln \frac{m_j^2}{\mu^2}-C_j\right]
\end{align}
where $C_j=\frac{5}{6}$ for the space-like polarizations of the gauge boson and $\frac{3}{2}$ for all other particles and where $\mu$ is the renormalization scale. 

Consider now the gauge-coupling dependence of each term. First, the contributions from the time-like gauge boson polarization and one ghost degree of freedom cancel completely. The contribution due to the space-like gauge boson is of order $e^4$ and is $\xi$-independent. The contributions from the Goldstone boson and the remaining ghost degree of freedom partially cancel against each other and produce a contribution of order $e^6$. 
%This can be seen by considering $m_{\chi}^4-m_{FP}^4$. 
Finally the Higgs boson produces a contribution of order $e^8$. We may now write the leading order potential as: 
\begin{align}
    V_{e^4}(\phi)=\frac{m^2}{2}\phi^2+\frac{\lambda}{4}\phi^4+\frac{3 m_A(\phi)^4}{64\pi^2}\left[\ln \frac{m_A^2}{\mu^2}-\frac{5}{6}\right]
\end{align}
where $m_A^2=2e^2\phi^2$ is the mass-squared of the {spacelike}
%transverse 
gauge boson. The important point to note here is that the effective potential is gauge-independent at this order. 

The resulting one-loop contributions to the $\mathcal{O}(e^6)$ effective potential yield a gauge-dependent terms
\begin{align}
    V_{e^6}^\xi(\phi)&=\frac{ m_\chi^4(\phi)}{64\pi^2}\left[\ln \frac{m_\chi^2(\phi)}{\mu^2}-\frac{3}{2}\right] \\
\nonumber
   & -\frac{ m_{FP}^4(\phi)}{64\pi^2}\left[\ln \frac{m_{FP}^2(\phi)}{\mu^2}-\frac{3}{2}\right]+\cdots\ \ \ ,
\end{align}
where $m_\chi$ is the  Goldstone boson mass,  $m_{FP}$ is the ghost mass, and the $+\cdots$ denote additional $\mathcal{O}(e^6)$ contributions arising from two-loop diagrams. As shown in \cite{Metaxas:1995ab}, it is possible to take into account a subset of these diagrams by replacing the tree level Goldstone boson mass by
\begin{align}
    m^2_{\chi}(\phi)&\to\frac{1}{2\phi}\pd{V_{e^4}}{\phi}+2\xi e^2\phi^2 \label{dressedmass}\\
    =\frac{1}{2}m^2&+\frac{3}{2}\lambda \phi^2+2\xi e^2\phi^2+\frac{e^4\phi^4}{8\pi^2}\left[3\ln \frac{2e^2\phi^2}{\mu^2}-1\right]
    \nonumber
\end{align}
We refer to this redefined $m_\chi$  as the dressed Goldstone boson mass. We discuss below that its inclusion is essential to obtain a gauge independent action at $\mathcal{O}(e^6)$. The remaining 2-loop diagrams sum to a gauge-independent contribution\cite{Metaxas:1995ab} (see also \cite{Andreassen:2016cvx}). For simplicity, we do not include these additional gauge-independent contributions further. 

To obtain a gauge-invariant effective action beyond $\mathcal{O}(e^4)$ one must also include higher order contributions to the wavefunction renormalization constant, $Z(\phi)$.  To do so, we proceed as in \cite{Garny:2012cg} and as discussed in Appendix \ref{app:wave}. 
Using the results listed there and using  $m^2_{\chi}=m^2_{FP}+\mathcal{O}(e^4)$ and $m^2_{FP}=\xi m^2_{A}=2\xi e^2\phi^2$, we obtain
\begin{align}
\nonumber
    Z(\phi)&=1+\frac{e^2}{16 \pi ^2} \left[\xi  \log \left(\frac{2e^2\phi^2 \xi }{\mu ^2}\right)+3 \log \left(\frac{2e^2\phi^2}{\mu ^2}\right)+\xi \right]\\
    & +\mathcal{O}(e^4)
 \label{wavefunctionrenormalization}
\end{align}

\noindent In the following sections, we will use these quantities calculated above to construct a gauge independent effective action.  

\section{The gauge-dependence problem}
\label{sec:problem}
Before discussing the procedure needed to obtain a gauge invariant tunneling rate, we first demonstrate the problem using the standard bounce solution and the new tunneling potential approaches when na\"ively implemented. {For purposes of illustration, we study a  concrete numerical example in parallel with the general, analytic discussion.}

\subsection{Standard bounce solution approach}
\label{sec:standardbounce}
{To be concrete, we take the following numerical values for the independent parameters: $e=0.07,\ \lambda=3.8\times 10^{-6},\ m^2=3.2\times 10^{-5}$ and $\mu=2$, where the units of the dimensionful parameters are arbitrary. Note that our choice introduces large logarithms in $V$, as needed to generate a perceptible barrier in the presence of the concavity of the tree-level potential\footnote{{Presumably one could re-sum these large logarithms using the renormalization group. We defer such a study to future work.}}. 
 %{\bf units are missing; also, doesn't this imply the logarithms are large?} \\ \\ 
%As a baseline, we first  take a naive approach by directly finding 
We obtain the bounce solution to the full equations of motion, using $V_{e^4}(\phi)+V_{e^6}^\xi(\phi)$, 
%corresponding to $V_{e^6}(\phi)$ using just 
the tree-level mass for $m_{\chi}$, and $Z(\phi)=1$.
%and ignoring the wavefunction renormalization.  
 Using the Mathematica package, FindBounce \cite{Guada:2020xnz}, we calculate the action over a range of values of $\xi$ in $[0,100]$ and find $\sim 13$\% variation, as indicated by the red dashed curve in Figure \ref{naive_espinosa}.
%\begin{figure}[htbp]
%    \centering
%    \includegraphics[scale=0.7]{naive.pdf}
%    \caption{The effective action as a function of the gauge dependence parameter, $\xi$. Here the bounce solution is calculated using $V_{e^6}$ while only using the tree-level mass for $m_{\chi}$. }
%    \label{naive_action}
%\end{figure}

\begin{figure}[htbp]
     \begin{center}
     \includegraphics[scale=0.65]{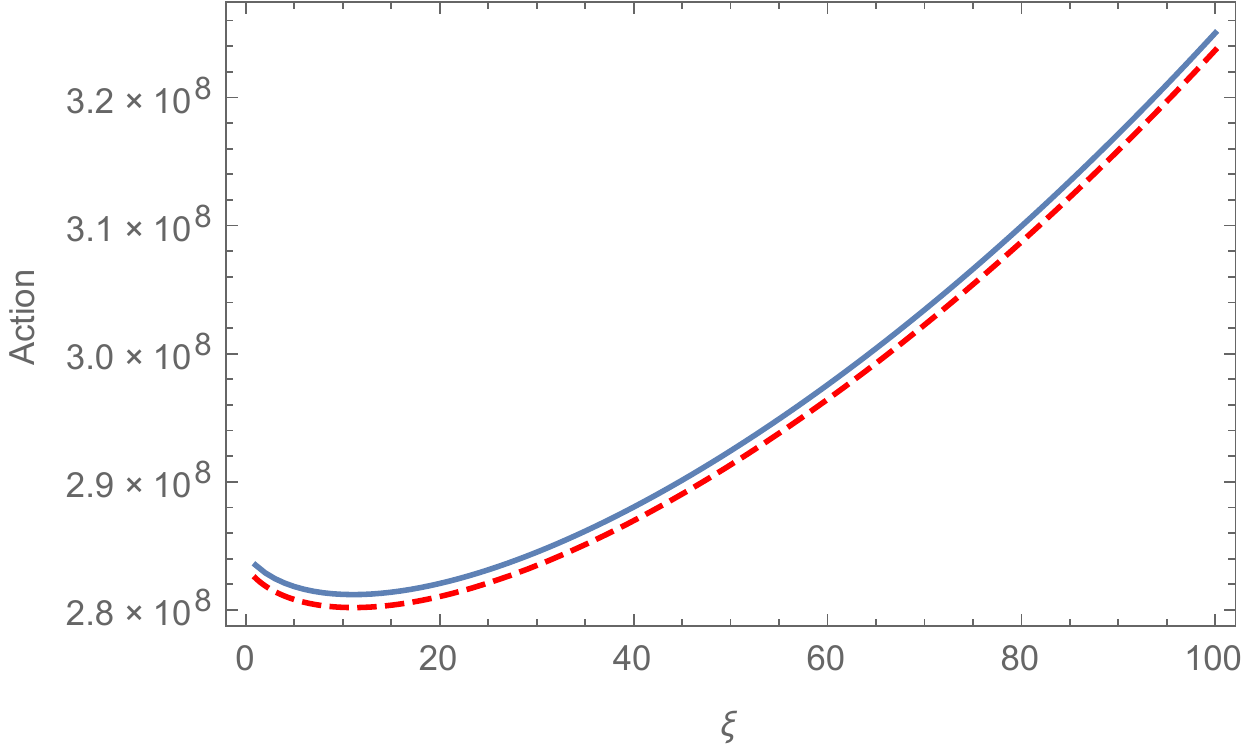}
     \end{center}
     \caption{The action evaluated as a function of  the gauge dependence parameter, $\xi$. 
 The red dashed curve gives the action computed using the standard bounce solution approach with the potential $V_{e^4}(\phi)+V_{e^6}^\xi(\phi)$, the tree-level Goldstone mass, and $Z(\phi)=1$.  The blue line gives the action calculated  following  the na\"ive tunneling potential method as outlined in Section \ref{sub:tunnel}. }
     \label{naive_espinosa}
 \end{figure}

The presence of $\xi$-dependence is not surprising, given the gauge dependence in $V_{e^6}^\xi(\phi)$. As we now show, the na\"ive application of the tunneling potential method inherits this $\xi$-dependence.
}

\subsection{Tunneling potential approach}
\label{sub:tunnel}

%In this section, we present an efficient approach to calculating a gauge-independent nucleation rate using the tunneling potential approach introduced by J.R. Espinosa in \cite{Espinosa:2018hue}. We will first review the key steps in this method and then show the modifications required to achieve gauge independence in the case of the Abelian Higgs theory discussed above. 

To review the logic of the tunneling potential introduced in Ref.~\cite{Espinosa:2018hue}, we start with the standard approach \cite{Coleman:1977py}, which entails first finding
%The standard approach to calculating the effective action \cite{Coleman:1977py} would be to first solve 
 the radially symmetric bounce solution in 4 dimensions  that satisfies the equation: 
\begin{align}
    \ddot{\phi}_b+\frac{3}{r}\dot{\phi}_b=\pd{V}{\phi}
\end{align}
with the boundary conditions: 
\begin{align}
    \dot{\phi}_b(0)=0,\ \ \phi_b(\infty)=\phi_+
\end{align}
where $\phi_+$ is the symmetric vacuum. One usually uses a shooting method to find the $\phi_0\equiv{\phi_b(0)}$ that satisfies the boundary conditions. 
In the method proposed in \cite{Espinosa:2018hue}, however, the  essential quantity is the tunneling potential, $V_t(\phi)$ given by 
\begin{align}
    V_t(\phi)\equiv V(\phi)-\frac{1}{2}\dot\phi_b^2 \label{Vtdef}
\end{align}
The idea behind introducing this quantity is to remove all explicit reference to the spacetime co-ordinates in the bounce equation and the effective action.
Immediately, one may use Eq.~(\ref{Vtdef}) to remove references to the derivative, $\dot{\phi}_b$, through 
\begin{align}
    \dot{\phi}_{b}=-\sqrt{2\left[V(\phi)-V_{t}(\phi)\right]}
\end{align}
Furthermore, one may rearrange the bounce equation to solve for the radial co-ordinate, $r$: 
\begin{equation}
    r=3 \sqrt{2\left(V-V_{t}\right) /\left(V_{t}^{\prime}\right)^{2}}
\end{equation}
Differentiating this expression with respect to $r$, one obtains:
\begin{align}
    \left(4 V_{t}^{\prime}-3 V^{\prime}\right) V_{t}^{\prime}=6\left(V_{t}-V\right) V_{t}^{\prime \prime} \label{espinosabounce}
\end{align} 
where primes denote the derivatives with respect to $\phi$. 

Eq.~(\ref{espinosabounce}) replaces the role of the bounce equation. One must now  find a $\phi_0$ and $V_t(\phi)$ that satsify \ref{espinosabounce} as well as the boundary conditions
\begin{align}
    V_t(\phi_+)=V(\phi_+),\ \ V_t(\phi_0)=V(\phi_0). \label{Espinosaboundary}
\end{align}
The Euclidean effective action can be written in terms of $V_t$ as:
\begin{align}
    S_E&=S_K+S_V\\
    S_K&=2\pi^2\int_{0}^\infty\frac{1}{2}\dot{\phi}_b^2 r^3dr =108 \pi^{2} \int_{\phi_{0}}^{\phi_+} \frac{\left(V-V_{t}\right)^{2}}{\left(V_{t}^{\prime}\right)^{3}} d \phi\\
    S_V&=2\pi^2\int_{0}^\infty V(\phi_b) r^3dr=108 \pi^{2} \int_{\phi_{0}}^{\phi_+} \frac{\left(V-V_{t}\right)}{\left(V_{t}^{\prime}\right)^{3}} V d \phi
\end{align}
One may now use the scaling relations, $S_{K}=2 S_{E}$ and $S_V=-S_E$ to write 
\begin{align}
    S_E=\frac{1}{2}S_K=54 \pi^{2} \int_{\phi_{0}}^{\phi_+}\frac{\left(V-V_{t}\right)^{2}}{\left(V_{t}^{\prime}\right)^{3}} d \phi \label{EspinosaS}
\end{align}
{As discussed in Appendix \ref{sec:scaling}, these relations apply only the presence of the tree-level form of the effective action and break down in the presence of higher order derivative terms.}

{At face value, solving Eq.~(\ref{espinosabounce}) is more challenging than solving the bounce equation.} However, the power of this method lies in the approximations one can make for $V_t$. As shown in \cite{Espinosa:2018hue}, one may choose for $V_t$, without much loss of accuracy, the quartic polynomial, $V_{t4}$ that satisfies 
%the boundary conditions, 
Eq.~(\ref{Espinosaboundary}) and solves Eq.~(\ref{espinosabounce}) only at the points, $\phi_0,\ \phi_+$, and $\phi_T$, {where the latter corresponds to } the top of the barrier. This approximation is given by: 
\begin{eqnarray}
    V_{t 4}(\phi)&=&V_{t 3}(\phi)+a_{4} \phi^{2}\left(\phi-\phi_{0}\right)^{2}\label{Vt4}\\
    V_{t 3}(\phi)&=&\frac{V_0}{\phi_0}\phi+\frac{1}{4 \phi_{0}^{2}}\left(3 \phi_{0} V_{0}^{\prime}-4 V_{0}\right)(\phi-\phi_0)\phi \\
    \nonumber
    &&+ \frac{1}{4 \phi_{0}^{3}}\left(3 \phi_{0} V_{0}^{\prime}-8 V_{0}\right) \phi(\phi-\phi_0)^2\ \ \ ,
\end{eqnarray}
where 
\begin{eqnarray}
    \nonumber
    a_{4}&=&\frac{1}{c}\left(a_{0 T}-\sqrt{a_{0 T}^{2}-c U_{t 3 T}}\right),\\
    \nonumber
     c &\equiv& 4 \phi_{T}^{2} \phi_{0 T}^{2}\left(\phi_{0}^{2}+2 \phi_{0 T} \phi_{T}\right)\\
    U_{t 3 T}& \equiv &
4\left(V_{t 3 T}^{\prime}\right)^{2}+6\left(V_{T}-V_{t 3 T}\right) V_{t 3 T}^{\prime \prime}\\
\nonumber
a_{0 T}&=&-6\left(V_{T}-V_{t 3 T}\right)\left(\phi_{0}^{2}-6 \phi_{0 T} \phi_{T}\right)\\
\nonumber
 &&-8 \phi_{T}\left(\phi_{0 T}-\phi_{T}\right) \phi_{0 T} V_{t 3 T}^{\prime}\\
 \nonumber
 &&+3 \phi_{T}^{2} \phi_{0 T}^{2} V_{t 3 T}^{\prime \prime}
\end{eqnarray}
with $V_0\equiv V(\phi_0)$, $\phi_{0 T} \equiv \phi_{0}-\phi_{T}$ and $V_{t 3 T} \equiv V_{t 3}\left(\phi_{T}\right)$.
We then obtain the best approximation of $S_E$ by finding the $\phi_0$ that minimises the integral in Eq.~(\ref{EspinosaS}). Hence, the difficult requirement of solving the bounce equation is replaced by solving a set of algebraic equations and minimising an integral. 
 
 %\subsection{Gauge independence and the tunneling potential approach\label{Espinosa_gauge}}
 
 {
We now show the gauge dependence of the above discussed tunneling potential approach and provide a prescription to eliminating this gauge dependence. 
 Suppose, firstly, that we define $V=V_{e^4}+V_{e^6}^\xi$ without using the dressed Goldstone mass. The na\"ive tunneling potential approach entails the following steps:
 \begin{enumerate}
     \item Calculate the quartic approximation to the tunneling potential, $V_{t4}$, using Eq.~(\ref{Vt4})
     \item Find the $\phi_0$ that minimizes the action as defined in Eq.~(\ref{EspinosaS})
     \item Evaluate this action for this value of $\phi_0$ 
 \end{enumerate}
 Following this approach, however, leads to a gauge-dependent result 
 that accurately reflects the $\xi$-dependence of the na\"ive bounce solution method, as one may observe by comparing the blue and red-dashed curves in Fig.~\ref{naive_espinosa}. It is clearly advantageous to translate the remedies for the gauge-dependence of the bounce action into the framework of the tunneling potential approach.
 }

\section{Gauge-independent nucleation rate using the derivative expansion\label{derivexpansion}}
\subsection{Overview}
{To set the stage, we review}
%We begin this section by reviewing 
the approach of Metaxas and Weinberg \cite{Metaxas:1995ab} in using the derivative expansion of the Nielsen identity to obtain a gauge independent nucleation rate.  The gauge dependence of the effective action is controlled by the Nielsen identity, 
\begin{eqnarray}
    \xi \frac{\partial S_\text{eff}}{\partial \xi} &=& -\int d^{4} x\Bigl\{
    \frac{\delta S_\text{eff}}{\delta A_{\mu}(x)} C_{A^{\mu}}(x)\\
    \nonumber
    &&+\frac{\delta S_\text{eff}}{\delta \phi(x)} C_{\phi}(x)+\frac{\delta S_\text{eff}}{\delta \phi^{*}(x)} C_{\phi^{*}}(x)\Bigr\}. 
\end{eqnarray}
This identity follows directly from the BRST invariance of the effective action. In computing $S_\text{eff}$ , one assumes the vector potential, $A_\mu$, is set to zero. Furthermore, performing this computation along the real axis, $\Phi=\Phi^*$, one obtains the identity, 
\begin{align}
    \xi \frac{\partial S_\text{eff}}{\partial \xi}=-\int d^{4} x\  C(x)\frac{\delta S_\text{eff}}{\delta \phi(x)} \label{Nielsen}
\end{align}
where 
\begin{align}
    C(x)&=\frac{i e}{2 \sqrt{2}} \int d^{4} y\Bigl\langle\bar{c}(x) \chi(x) c(y)\times
\\    
\nonumber  
   &\Bigl(\partial_{\mu} A^{\mu}(y)  
   -\sqrt{2} e \xi \bar{\phi} \chi(y)\Bigr)\Bigr\rangle. 
\end{align}

Now, one may perform a gradient expansion of both $S_\text{eff}$ and $C(x)$ as:
\begin{align}
    S_\text{eff}&=\int d^{4} x\left[V_\text{eff}(\phi)+\frac{1}{2}Z(\phi) \left(\partial_{\mu} \phi\right)^{2}+\mathcal{O}\left(\partial^{4}\right)\right]\label{gradient1}\\
    C(x)&=C_{0}(\phi)+D(\phi)\left(\partial_{\mu} \phi\right)^{2}-\partial^{\mu}\left[\tilde{D}(\phi) \partial_{\mu} \phi\right]+\mathcal{O}\left(\partial^{4}\right)\label{gradient2}\\
    \frac{\delta S_\text{eff}}{\delta \phi(x)}&=\frac{\partial V_{\mathrm{eff}}(\phi)}{\partial \phi}+\frac{1}{2} \frac{\partial Z}{\partial \phi}\left(\partial_{\mu} \phi\right)^{2}-\partial_{\mu}\left[Z(\phi) \partial_{\mu} \phi\right]+\mathcal{O}\left(\partial^{4}\right)\label{gradient3}
\end{align}
Note that the total derivative term involving $\tilde{D}(\phi)$ was not included in the derivative expansion of $C(x)$ in \cite{Metaxas:1995ab} but could contribute to this identity as pointed out in \cite{Garny:2012cg}. 
Expanding Eq.~(\ref{Nielsen}) using Eqs.~(\ref{gradient1}-\ref{gradient3}), we obtain the Nielsen identity at {orders $\partial^0$ and $\partial^2$, respectively, as}
\begin{align}
\xi \frac{\partial V_{\text{eff}}}{\partial \xi}&=-C_0 \frac{\partial V_{\text{eff}}}{\partial \phi}\label{p0Nielsen}\\
\xi \frac{\partial Z}{\partial \xi}&=-C_0 \frac{\partial Z}{\partial \phi}-2 D \frac{\partial V_{\mathrm{eff}}}{\partial \phi}-2 \tilde{D} \frac{\partial^2 V_{\mathrm{eff}}}{\partial \phi^2}-2 Z \frac{\partial C_0}{\partial \phi}.\label{p2Nielsen}
\end{align} 
For a detailed derivation of the Nielsen identity, one may refer to \cite{Nielsen:1975fs,Fukuda:1975di,Aitchison:1983ns,Metaxas:1995ab,Garny:2012cg}.

{The approach of Ref.~\cite{Metaxas:1995ab} for obtaining a gauge-independent nucleation rate using the aforementioned derivative expansion proceeds as follows. 
%As discussed in the previous section, one may adopt a power counting, $\lambda\sim e^4$, 
Using the power counting introduced in Section \ref{model} leads to the gauge coupling expansion}
\begin{align}
    V_\text{eff}&=V_{e^4}+ V_{e^6}+\mathcal{O}\left(e^8\right)\\
     Z&=1+Z_{e^2}+\mathcal{O}\left(e^4\right)
\end{align}
As we  show explicitly below, the leading order contribution to $C_0$ is of order $e^2$. As pointed out in \cite{Metaxas:1995ab}, the leading order contributions to $D$ and $\tilde{D}$ are of order $e^0$. Thus, the leading and sub-leading order contributions to Eqs. (\ref{p0Nielsen}) and (\ref{p2Nielsen}) are given by: 
\begin{align}
    \xi\pd{V_{e^4}}{\xi}&=0\label{Ve4Nielsen}\\
    \xi\pd{V_{e^6}}{\xi}&=-C_{e^2}\pd{V_{e^4}}{\phi}\label{Ve6Nielsen}\\
    \xi\pd{Z_{e^2}}{\xi}&=-2\pd{C_{e^2}}{\phi}\label{Ze2Nielsen}
\end{align}
One may further write the leading and subleading order terms of $B$ in Eq. (\ref{nucleationdef})  as
\begin{align}
    B_\text{eff}&=B_0+B_1\\
    B_0&=\int d^{4} x\left(\frac{1}{2}\left(\partial_{\mu} \phi_b\right)^{2}+V_{e^4}(\phi_b)\right)\label{B0}\\
    B_1&=\int d^{4} x\left(\frac{1}{2}Z_{e^2}\left(\partial_{\mu} \phi_b\right)^{2}+V_{e^6}(\phi_b)\right)\label{B1}
\end{align}
As shown in \cite{Metaxas:1995ab}, Eqs. (\ref{Ve4Nielsen}-\ref{Ze2Nielsen}) ensure that $S_0$ and $S_1$ are gauge independent when evaluated at the leading order bounce solution, satisfying the equation:
\begin{align}
    \Box \phi_b=\pd{V_{e^4}}{\phi}.\label{Bounce}
\end{align}
The gauge independence of $B_0$ can be seen immediately as Eq. (\ref{Ve4Nielsen}) directly implies the gauge independence of $V_{e^4}$ and consequently, the bounce solution. Using Eqs. (\ref{Ve6Nielsen}) and (\ref{Ze2Nielsen}), we see that 
    
\begin{align}
    \xi\pd{B_1}{\xi}=-\int d^{4} x\left[\pd{C_{e^2}}{\phi}\left(\partial_{\mu} \phi_b\right)^{2}+C_{e^2}(\phi_b)\pd{V_{e^4}(\phi_b)}{\phi}\right]
\end{align}
Noting that $\pd{C_{e^2}}{\phi}\partial_\mu\phi=\partial_\mu C_{e^2}$ and integrating by parts, one obtains
\begin{align}
    \xi\pd{B_1}{\xi}=\int d^{4} x\  C_{e^2}(\phi_b)\left[\Box\phi_b-\pd{V_{e^4}(\phi_b)}{\phi}\right]\label{gaugeindependence}
\end{align}
Hence, the bounce equation, Eq. (\ref{Bounce}), guarantees the gauge independence of $B_1$. 

\subsection{Explicit verification of the Nielsen identities}
{For the purposes of this paper, it is useful to explicitly verify Eqs. (\ref{Ve6Nielsen}) and (\ref{Ze2Nielsen}) -- the key ingredients in maintaining gauge invariance of $B_1$. The Nielsen functional $C_{e2}$ receives a sole contribution from the diagram in Fig.~\ref{Cfunctional}:}
\begin{align}
    C_{e^2}=\frac{-ie^2\xi\phi}{2}\int \frac{d^4k}{(2\pi)^4}\frac{1}{\left(k^2-m_{\chi}^2\right)\left(k^2-m^2_{FP}\right)}.
\end{align}

%We may now directly verify Equation \ref{Ve6Nielsen} without evaluating any integrals. 
Recalling that 
\begin{equation}
\pd{m_{\chi}^2}{\xi}=\pd{m_{FP}^2}{\xi}=2e^2\phi^2\ \ \ ,
\end{equation}
and
\begin{equation}
V_{e^6}^\xi = \frac{i}{2}\int \frac{d^4k}{(2\pi)^4}
\left[\ln\left(k^2-m_{FP}^2\right) -\ln\left(k^2-m_\chi^2\right)\right]
\ee
 we see that 
% \begin{widetext}
\begin{eqnarray}
\nonumber
    \xi\pd{V_{e^6}}{\xi}&=&\xi\pd{}{\xi}\int \frac{d^4k}{(2\pi)^4}\Bigl[-\frac{i}{2}\ln\left(k^2-m_{\chi}^2\right) +\frac{i}{2}\ln\left(k^2-m_{FP}^2\right)\Bigr]\\
    \nonumber
        &=& i\xi e^2\phi^2 \left(m^2_{\chi}-m^2_{FP}\right)\int \frac{d^4k}{(2\pi)^4}\frac{1}{\left(k^2-m_{\chi}^2\right)\left(k^2-m^2_{FP}\right)}
    =-C_{e^2}\pd{V_{e^4}}{\phi}\ \ \ ,
\end{eqnarray}
%\end{widetext}
{as required by Eq. (\ref{Ve6Nielsen}). Importantly, to obtain the last equality,  we have used  the dressed Goldstone mass squared $m_\chi^2$ given in Eq. (\ref{dressedmass}). Use of the tree-level mass would not allow the the Nielsen identiy at this order to be satisfied.}

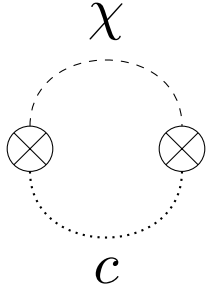
\begin{figure}[htbp]
    \centering
   \feynmandiagram [scale=2,transform shape][layered layout, horizontal=b to c] {
 b [crossed dot,black]
-- [scalar, half left,edge label = $\chi$] c [crossed dot]
-- [ghost, half left,edge label = $c$] b
};
    \caption{The diagram contributing to $C_{e^2}$}
    \label{Cfunctional}
\end{figure}

To verify Eq. (\ref{Ze2Nielsen}), the most convenient approach is to evaluate the correlator integral. Using the $\overline{MS}$ renormalization scheme, one obtains 
\begin{align}
    C_{e^2}=e^2 \xi  \phi\frac{m_{FP}^2 \left[\log \left(\frac{m^2_{FP}}{\mu ^2}\right)-1\right]-m^2_\chi\left[ \log \left(\frac{m^2_\chi}{\mu ^2}\right)-1\right]}{32 \pi ^2 (m^2_\chi-m^2_{FP})}
\end{align}
Keeping only terms of order $e^2$, one obtains
\begin{align}
    C_{e^2}=-\frac{e^2 \xi \phi}{32 \pi ^2}\log \left(\frac{2 e^2 \xi  \phi ^2}{\mu ^2}\right)+\mathcal{O}\left(e^4\right)\label{Ce2}
\end{align}
{Using Eqs. (\ref{Ce2}) and (\ref{wavefunctionrenormalization}) leads immediately 
to verification of Eq. (\ref{Ze2Nielsen}).}

\subsection{Gauge invariance with the derivative expansion\label{numericderiv}}

\begin{comment}
In this section, we take a more quantitative approach at studying the gauge dependence of the effective action. For this analysis, we chose the parameters, $e=0.07,\ \lambda=3.8\times 10^{-6},\ m^2=3.2\times 10^{-5}$ and $\mu=2$.\\ \\ As a baseline, we first  take a naive approach by directly finding the bounce solution corresponding to $V_{e^6}(\phi)$ using just the tree-level mass for $m_{\chi}$ and ignoting the wavefunction renormalization.  Using the Mathematica package, FindBounce \cite{Guada:2020xnz}, we calculated the action over a range of values of $\xi$ in $[0,100]$ and found $\sim 13$\% variation. The results may be seen in Figure \ref{naive_action}.
\begin{figure}[htbp]
    \centering
    \includegraphics[scale=0.7]{naive.pdf}
    \caption{The effective action as a function of the gauge dependence parameter, $\xi$. Here the bounce solution is calculated using $V_{e^6}$ while only using the tree-level mass for $m_{\chi}$. }
    \label{naive_action}
\end{figure}
\end{comment}
{We  now implement carefully the gradient expansion of Ref.~\cite{Metaxas:1995ab} discussed in the previous sections. The three key modifications of the na\"ive computation  are: }
\begin{enumerate}
    \item Using the bounce solution corresponding to $V_{e^4}$ as seen in Eq. (\ref{Bounce})
    \item Using a dressed mass for $m_{\chi}$ when evaluating the action. This is essential for the Nielsen identity at $\mathcal{O}(\partial^0)$, Eq. (\ref{Ve6Nielsen}), to be satisfied. 
    \item Including the wavefunction renormalization, $Z_{e^2}$. As shown in Eq. (\ref{gaugeindependence}), this is required to cancel the gauge dependence in the effective potential. 
\end{enumerate}

{Fig. \ref{gradient_action_findbounce} illustrates the impact of each modification, using the same numerical inputs for the parameters as employed in obtaining Fig.~\ref{naive_espinosa}. }Implementing the first modification slightly reduces the gauge dependence as the bounce solution itself is no longer gauge-dependent. Numerically calculating the action with this modification in the interval, $\xi\in[0,100]$, we found a variation of 11\%. Using the dressed mass for $m_{\chi}$ significantly suppresses the gauge-dependence of the effective potential as this causes the potential evaluated  at the  radiatively generated minimum of $V_{e^4}$ to be gauge independent. We found a significantly lower variation of $0.9\%$. Finally, incorporating the wavefunction renormalization as well, one obtains a variation of $\lesssim 0.001\%$. 
%These are seen in Figure \ref{gradient_action_findbounce}. 
{This residual gauge dependence arises because the full analytic expression for $V_{e^6}^\xi$ -- which we use for computational convenience -- contains gauge-dependent higher order terms in $e$. In principle, had we truncated $V_{e^6}^\xi$ at $\mathcal{O}(e^6)$ then $B_1$ would be exactly $\xi$-independent.}

\begin{figure}
    \centering
    \includegraphics[scale=0.65]{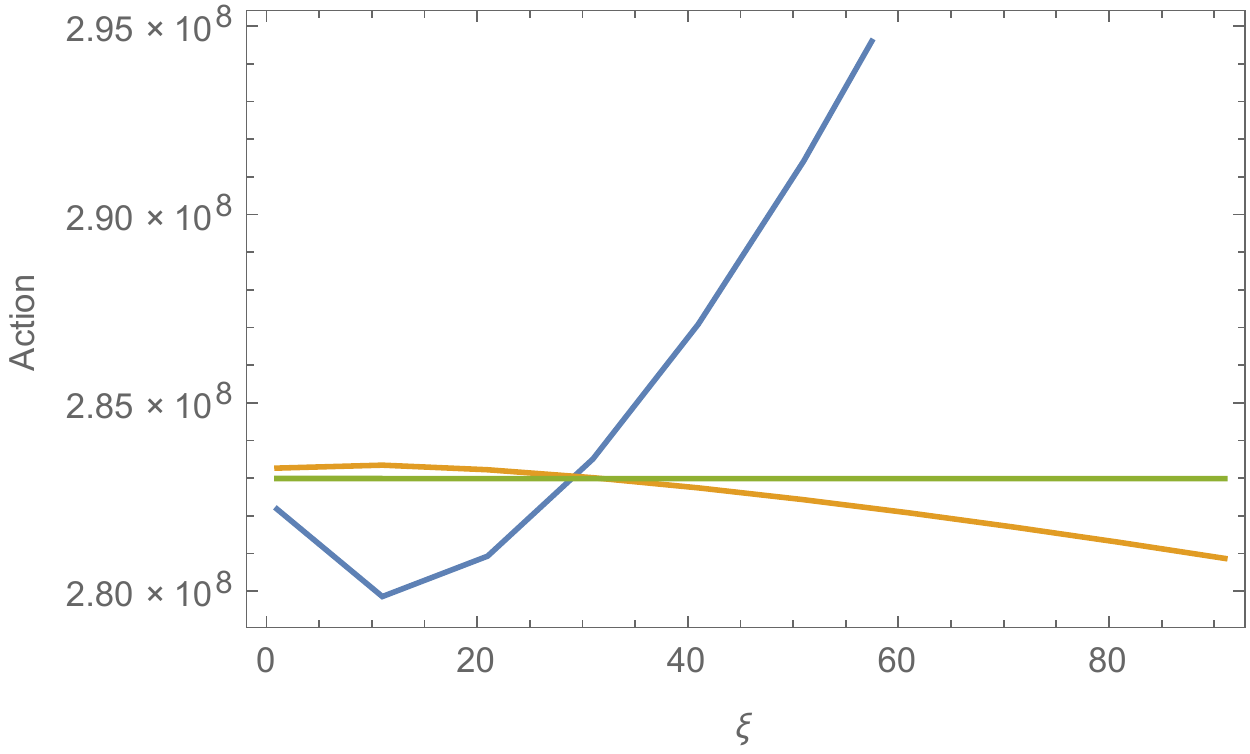}
    \caption{The gauge dependence of the effective action using the modifications proposed by the derivative expansion. The blue curve corresponds to the first modification. The orange curve corresponds to both the first two modifications. The green curve depicts the action incorporating all three modifications proposed in \cite{Metaxas:1995ab}}
    \label{gradient_action_findbounce}
\end{figure}

\section{Gauge-independent nucleation rate using tunneling
potentials\label{tunnellingpotentialapproach}}

 \subsection{Gauge independence and the tunneling potential approach\label{Espinosa_gauge}}
 \color{black}
 {In this section, we show how to translate the derivative expansion approach into the tunneling potential framework in order to eliminate the gauge-dependence exhibited by the blue curve in Fig. \ref{naive_espinosa}. }

 \begin{comment}
 
 Suppose, firstly, that we define $V=V_{e^4}+V^\xi_{e^6}$ without using the dressed Goldstone mass. The na\"ive tunneling potential approach would be to follow the following steps:
 \begin{enumerate}
     \item Calculate the quartic approximation to the tunneling potential,$V_{t4}$, using Eq. (\ref{Vt4})
     \item Find the $\phi_0$ that minimizes the action as defined in Eq. \ref{EspinosaS}
     \item Evaluate this action for this value of $\phi_0$ 
 \end{enumerate}
 Following this approach, however, leads to a gauge dependent result similar to the the findings in Figure \ref{naive_action}. This may be seen in Figure \ref{naive_espinosa} where we have compared the results of the tunneling potential method with that of the bounce action approach. It is apparent that the gauge dependence is consistent between the two methods. 
 
 \end{comment}
 
 \begin{comment}
 
 \begin{figure}[htbp]
     \centering
     \includegraphics{naiveespinosa.pdf}
     \caption{The action evaluated as a function of  the gauge dependence parameter, $\xi$. The blue line is the action calculated naively following  the tunneling potential method while the red dashed line is the same as the curve in Figure \ref{naive_action}. }
     \label{naive_espinosa}
 \end{figure}

\end{comment}

{We start by adopting the power-counting in $e$ given in Section \ref{model}  and perform the same three modifications as listed in Section \ref{numericderiv}. The key steps are as follows.

\noindent {\it Step 1}: Calculate the leading order action, $B_0$, using the na\"ive tunneling potential method. Specifically, we compute the quartic approximation of the tunneling potential using Eq. (\ref{Vt4}) with $V_0=V_{e^4}(\phi_0)$ and $\phi_T$ being the maximum of $V_{e^4}$. We then find the $\phi_0$ that minimizes the integral,  
 \begin{align}
 B_{0,\text{approx}}= 54 \pi^{2} \int_{\phi_{0}}^{\phi_+} \frac{\left(V_{e^4}-V_{t4}\right)^{2}}{\left(V_{t4}^{\prime}\right)^{3}} d \phi
\label{B0approx} \end{align}
 and compute the integral using this value of $\phi_0$. Note that so far, no gauge dependence has been introduced as $V_{e^4}$ is gauge-independent. 
 
\noindent {\it Step 2}: Calculate $B_1$ starting with the expression:}
% \begin{widetext}
 \begin{align}
       B_{1,\text{approx}}&= S_{K,1}+S_{V,1}\\
     &= 108 \pi^{2}\left( \int_{\phi_{0}}^{\phi_+} \left[\frac{\left(V_{e^4}-V_{t4}\right)^{2}}{\left(V_{t4}^{\prime}\right)^{3}}\right]_{e^4} Z_{e^2}(\phi)\ d \phi+ \int_{\phi_{0}}^{\phi_+}\left[\frac{\left(V_{e^4}-V_{t4}\right)}{\left(V_{t4}^{\prime}\right)^{3}}\right]_{e^4} V_{e^6}(\phi)\ d \phi\right)\label{B1approx}
 \end{align}
% \end{widetext}
 where the quantities in the square brackets and $\phi_0$ are the ones obtained from minimizing $B_{0,\text{approx}}$. This expression satisfies the three modifications listed in Section \ref{numericderiv} in the following way:
 \begin{itemize}
     \item {\bf The leading order bounce solution:} The role of the bounce solution is replaced by $V_t$ and $\phi_0$ in the tunneling potential approach {as indicated in Step 1}. Hence, evaluating the quantities in the square brackets using the $\phi_0$ found by minimising $B_{0,\text{approx}}$ is the equivalent to evaluating the action using the leading order bounce solution. 
     \item {\bf Using the dressed Goldstone propagator:} This is achieved by introducing the $S_{V,1}$  term with the dressed propagator included in $V_{e^6}$. {Note that with our choice of $V_t$, the scaling relation $S_K=2S_E$ holds only at $\mathcal{O}(e^4)$. Hence, we must explicitly compute the kinetic and potential contributions to $B_1$.  Satisfying Eq. (\ref{Ve6Nielsen}) then requires the use of the dressed Goldstone propagator, as discussed above.}
     
%The reason for including this term is that the choice of $V_t$ does not minimize $S_{K,1}$ and hence, we may no longer rely on the scaling relation. \mrm{\bf I don't quite understand this statement, how use of the dressed mass is connected to breaking of the scaling relations.}
     \item {\bf Including the wavefunction renormalization:} This is achieved by the inclusion of $Z_{e^2}(\phi)$ in $S_{K,1}$. 
 \end{itemize}
 
{We note that  should one  have in hand an exact $V_t$ that satisfies  Eq.~(\ref{espinosabounce}), the above steps would result in a gauge independent calculation of $B_1$. However, there is an additional numerical concern that arises due to the use of the quartic approximation for the tunneling potential. This will be discussed in the next section. }

 \subsection{Gauge dependence of the tunneling potential approach: a numerical study}
 
In this section, we numerically investigate the use of the procedure outlined in the previous section. Using the values for the parameters as in Section \ref{sec:standardbounce},  we begin by investigating the accuracy of the calculation of $B_0$. Computing the effective action using Eq.~(\ref{EspinosaS}) and minimizing with respect to $\phi_0$, one obtains an effective action that agrees with the results of the FindBounce package with an error of approximately 0.3\%. This can be seen in Fig.~\ref{espinosae4approx}. 
 \begin{figure}[htbp]
     \centering
     \includegraphics[scale=0.7]{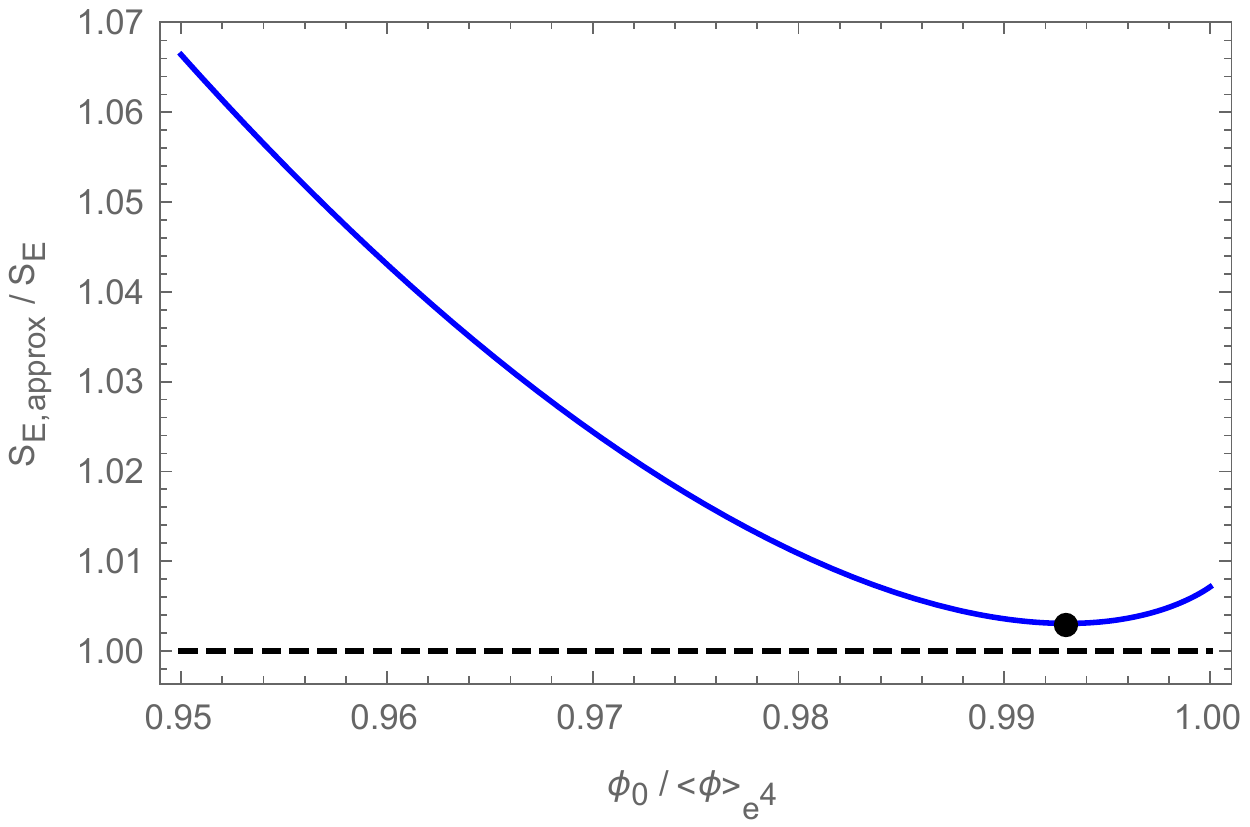}
     \caption{The estimate of the tunneling action by calculating the integral in Eq.~(\ref{EspinosaS} )using the quartic approximation in Eq.~(\ref{Vt4}). {We normalize $\phi_0$  by the broken minimum of $V_{e^4}$ and the action to the action computed using Findbounce}. The black dot represents the minimum of the integral in Eq.~(\ref{B0approx}).}
     \label{espinosae4approx}
 \end{figure}
 
 Similarly, calculating $B_1$ using Eq.~(\ref{B1approx}), one obtains Fig.~\ref{Espinosa_without_scaling}. We find that there is an error of $0.4\%-0.5\%$ when compared to the tunneling rate calculated in Section \ref{numericderiv} and  a gauge dependence of order 0.1\%. 
 \begin{figure}
     \centering
     \includegraphics[scale=0.7]{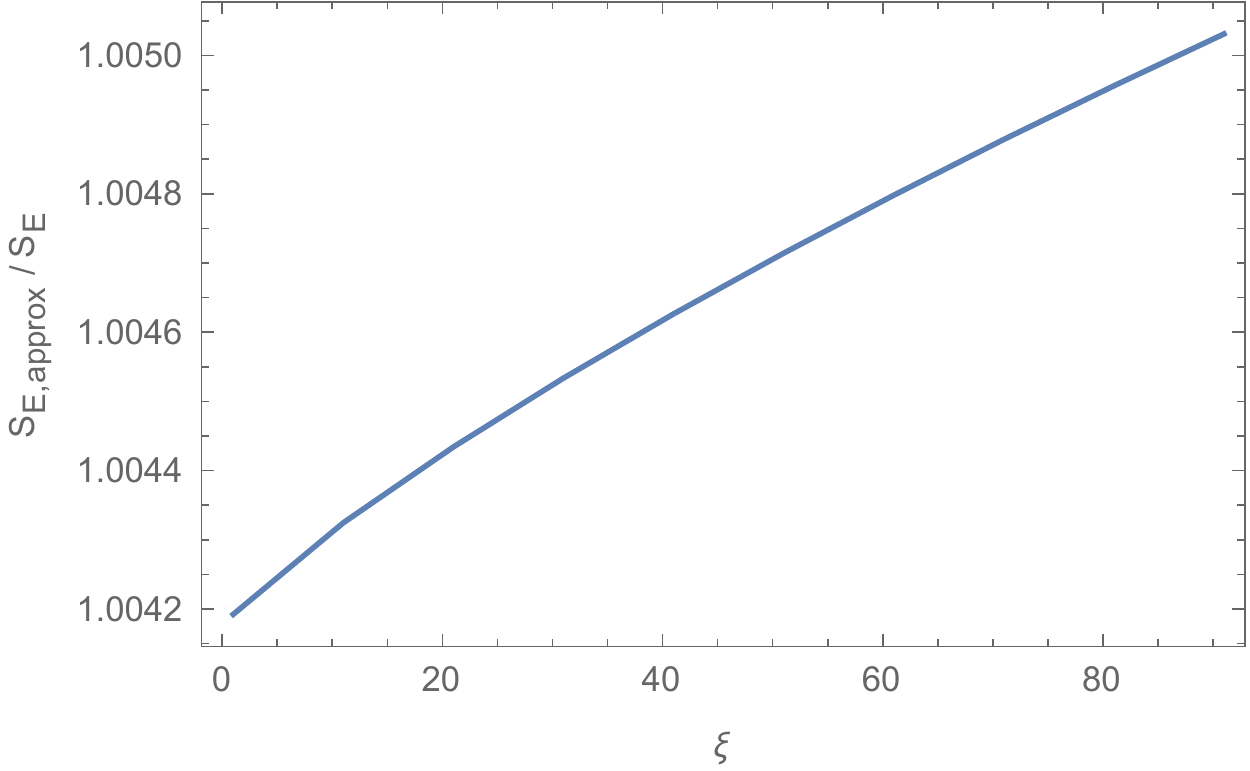}
     \caption{The action calculated using the procedure outlined in Section \ref{Espinosa_gauge}. These are normalized using the corresponding values found using the bounce solution in Section \ref{numericderiv}. }
     \label{Espinosa_without_scaling}
 \end{figure}. 
 \color{black}
 
One can further {reduce the gauge-dependence} of this approximation by looking more deeply into the scaling relation, $S_K=-\frac{1}{2}S_V$. Suppose one uses the value of $\phi_0$ that minimizes the integral Eq.~(\ref{B0approx}) and calculates the action using $S_E=-S_{V}$, we find a relatively large error of approximately 15$\%$. The source of this error is {the sensitivity of} $S_V$  to small variations in $\phi_0$. In contrast, $S_K$ is quite robust to these variations, as this $\phi_0$ is a stationary point. This is evident in Fig.~ \ref{espinosae4pot} ,where we see  that $\frac{1}{2}S_K$ is almost constant while $-S_V$  varies dramatically {with $\phi_0$}.   We note that the large error in $S_V$ is  artifact of using the quartic approximation. In the case of exactly solvable tunneling potentials, the minimum of $\frac{1}{2}S_K$ exactly corresponds the point where the scaling relation is satisfied.  

{To further reduce the $\xi$-dependence for a general potential,} we propose, instead, that one should use the $\phi_0$ that satisfies $\frac{1}{2}S_K=-S_V$ (the point of intersection in Fig.~\ref{espinosae4pot}) to compute $B_0$ and $B_1$. 
 Calculating $B_0$ in this way, we find an error of approximately $0.32\%$ with respect to the action calculated using FindBounce. This is only a slight increase in the error for $B_0$ when compared to minimizing the integral in Eq.~(\ref{B0approx}). 
 \begin{figure}[htbp]
     \centering
     \includegraphics[scale=0.7]{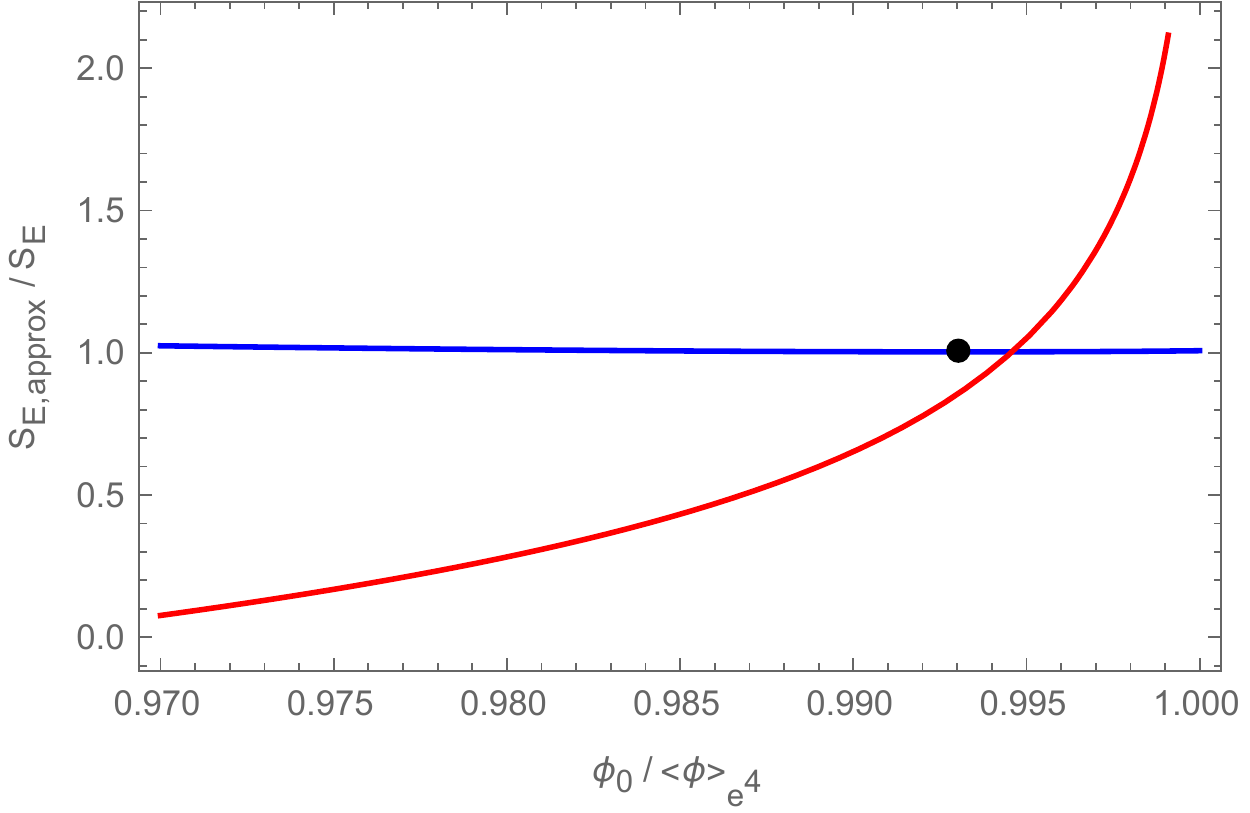}
     \caption{The actions computed using $S_E=\frac{1}{2}S_K$(blue) and $S_E=-S_V$ (red). The point of intersection is the point where the scaling relations are satisfied while the black dot represents the minimum of $S_K$.}
     \label{espinosae4pot}
 \end{figure}

{Following the foregoing steps and } evaluating $B_{0,\text{approx}}+B_{1,\text{approx}}$ for $\xi\in[1,100]$,  we obtain Fig.~\ref{espinosafinal}. 
{Comparing with Fig.~\ref{Espinosa_without_scaling}, we see a slightly larger error for small $\xi$} by using this value of $\phi_0$ {rather than the one which minimizes the integral Eq.~(\ref{B0approx}). However, }  the  gauge dependence when using this approach is of order $0.05\%$ in this range. This makes sense as the tunneling potential now satisfies the scaling relation.  The residual gauge dependence {occurs because} the $\xi$-dependence of $B_1$ is only zero if the bounce equation is satisfied {exactly, as seen in Eq.~ (\ref{gaugeindependence}), whereas} the quartic approximation being used for $V_{t}$ only satisfies the bounce equation at the three points, $\phi_+,\phi_0$ and $\phi_T$. Nevertheless, the efficiency and convenience of this approach justifies its use when compared to solving the full bounce equation.  
\begin{figure}[htbp]
    \centering
    \includegraphics[scale=0.7]{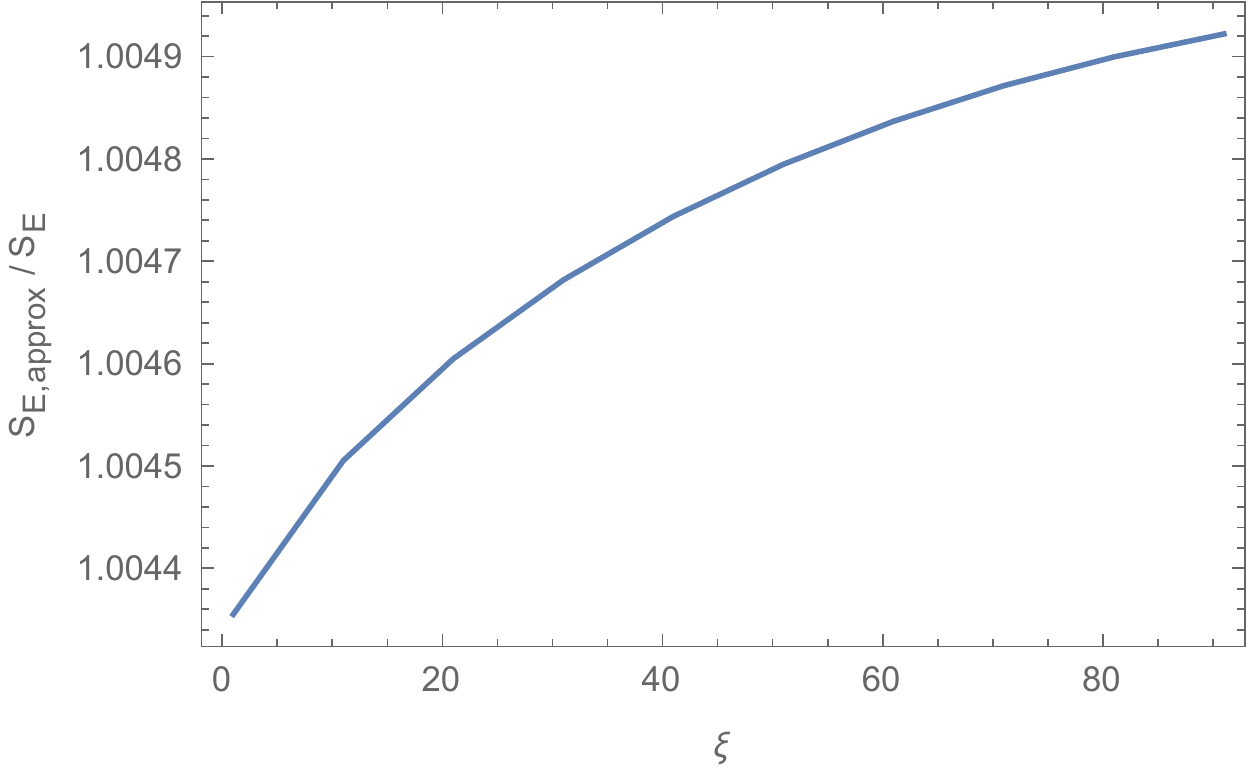}
    \caption{This shows the approximate effective action as a function of $\xi$ using the tunneling potential approach  as defined in Eqs.~(\ref{B0approx}) and (\ref{B1approx}). These are normalized using the corresponding values found using the bounce solution in Section \ref{numericderiv}.  }
    \label{espinosafinal}
\end{figure}

\section{Summary\label{Summary}}

{Computing the nucleation rate is a key step in exploring the possibility of  first order phase transitions in the early universe. The problem can be particularly challenging for scenarios involving multiple scalar fields, such as Higgs portal models that may lead to a first order electroweak phase transition. The recent development of the tunneling potential method provides a computationally efficient and numerically accurate approach for addressing this challenge. It is well-known that in conventional approaches, care must be taken to ensure that the resulting nucleation rate is gauge-invariant as required for a physical observable. We are, thus, motivated to investigate gauge-invariance in the tunneling potential framework.  }

{In this paper, we have shown a way to calculate a gauge independent nucleation rate at $T=0$ using tunneling potentials in the Abelian Higgs theory with a radiatively induced barrier. We first reviewed how using a derivative expansion and a power counting of $\lambda\sim e^4$ allows one to obtain a gauge-independent nucleation rate. We then showed how the tunneling potential approach can be modified to include these expansions consistently. In particular, one must compute the tunneling potential, $V_t$, and the initial field value, $\phi_0$, at leading order before using these to calculate higher order corrections. Inclusion of the dressed Goldstone mass and the $\mathcal{O}(e^2)$ contribution to the wavefunction renormalization is essential. We also showed that when using the quartic approximation to the tunneling potential, the scaling relation between the kinetic and potential parts of the action must be directly imposed due to the numerical sensitivity of $S_V$ to $\phi_0$. Application of these ideas to the nucleation rate at $T>0$ will appear in a forthcoming publication.}

\acknowledgments
We thank J. L\"ofgren and T.V.I. Tenkanen for helpful discussions and J. L\"ofgren for a careful reading of this manuscript. This work was supported in part under National 
Natural Science Foundation of China grant No. 19Z103010239. 

\appendix

\section{Wavefunction Renormalization}
\label{app:wave}
This may be calculated in an analogous way to what is done in \cite{Garny:2012cg} by finding the  leading order $\mathcal{O}(p^2)$ contributions to the diagrams in Figure \ref{Zdiagrams}. 
Writing the kinetic part of the effective action as: 
\begin{align}
    S_K=\int d^4x \frac{1}{2}Z(\phi)\left(\partial_\mu\phi\right)^2
\end{align}
we find that $Z(\phi)$
%[MICHAEL: Here is the expression I calculated by reverse engineering Garny and Konstandin. However, when Tuomas tried to calculate it from scratch, he found some factors of 2 differences in some of these. I am pretty confident with the final answer though as it satisfies the Nielsen identity. I will elaborate on my calculations in the notes I will write so it will be easier to check.] 
is given by:
\begin{align}
\nonumber
    Z(\phi)&=1+Z_a(\phi)+Z_b(\phi)+Z_c(\phi)+Z_d(\phi)+\mathcal{O}(e^4)\\
    \nonumber
    Z_a(\phi)&=\frac{5e^2}{32\pi^2}+\frac{11\xi e^2}{96\pi^2}-\frac{3e^2}{16\pi^2} \frac{m_{FP}^2\ln\left(\frac{m^2_{FP}}{m_A^2}\right)}{m_{FP}^2-m_A^2}\\
    \nonumber
    Z_b(\phi)&=\frac{e^2\xi}{48\pi^2}\left(\frac{m_{FP}^2}{m_{\chi}^2}\right)\\
    \nonumber
    Z_c(\phi)&=-\frac{e^2\xi}{24\pi^2}\\
    \nonumber
    Z_d(\phi)&=\\
    \nonumber
    \frac{e^2 \xi}{16 \pi ^2} & \frac{  m^2_{FP}\left( \log \left(\frac{m^2_{FP}}{\mu ^2}\right)-\frac{3}{2}\right) -m^2_\chi \left(\log \left(\frac{m^2_\chi}{\mu ^2}\right)-\frac{3}{2}\right)}{ (m^2_{FP}-m^2_\chi)}\\
    \nonumber
    +\frac{3e^2}{16\pi^2} & \frac{  m^2_{A}\left( \log \left(\frac{m^2_{A}}{\mu ^2}\right)-\frac{5}{6}\right) -m^2_\chi \left(\log \left(\frac{m^2_\chi}{\mu ^2}\right)-\frac{5}{6}\right)}{ (m^2_{A}-m^2_\chi)}
\end{align}

\begin{figure}[htbp]
   \begin{subfigure}[c]{0.475\textwidth}
   \feynmandiagram [scale=1.3,transform shape][layered layout, horizontal=b to c] {a [] -- [scalar,edge label=$h$]
 b []
-- [boson, half left,edge label = $A_\mu$] c [] 
-- [boson, half left,edge label = $A_\mu$] b,
c -- [scalar,edge label=$h$] d
};
\subcaption{}
\end{subfigure}
\hfill
  \begin{subfigure}[c]{0.475\textwidth}
   \feynmandiagram [scale=1.3,transform shape][layered layout, horizontal=b to c] {a [] -- [scalar,edge label=$h$]
 b []
-- [scalar, half left,edge label = $\chi$] c [] 
-- [scalar, half left,edge label = $\chi$] b,
c -- [scalar,edge label=$h$] d
};
\subcaption{}
\end{subfigure}
\vskip\baselineskip
   \begin{subfigure}[c]{0.475\textwidth}
   \feynmandiagram [scale=1.3,transform shape][layered layout, horizontal=b to c] {a [] -- [scalar,edge label=$h$]
 b []
-- [ghost, half left,edge label = $c$] c [] 
-- [ghost, half left,edge label = $\overline{c}$] b,
c -- [scalar,edge label=$h$] d
};
\subcaption{}
\end{subfigure}
\hfill
  \begin{subfigure}[c]{0.475\textwidth}
   \feynmandiagram [scale=1.3,transform shape][layered layout, horizontal=b to c] {a [] -- [scalar,edge label=$h$]
 b []
-- [boson, half left,edge label = $A_\mu$] c [] 
-- [scalar, half left,edge label = $\chi$] b,
c -- [scalar,edge label=$h$] d
};
\subcaption{}
\end{subfigure}
    \caption{The $\mathcal{O}(p^2)$ terms in these diagrams form the   leading order contributions to $Z(\phi)$}
    \label{Zdiagrams}
\end{figure}
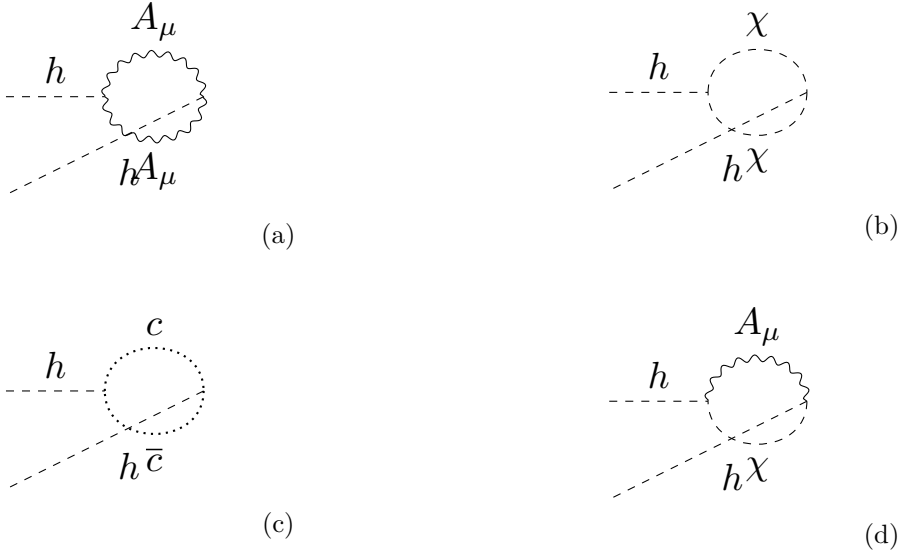

\section{Scaling relations}
\label{sec:scaling}

{The scaling relations $S_E=S_K/2=-S_V$ follow from the logic given in Ref.~\cite{Derrick:1964ww}. Recalling that the leading order Euclidean action  $S_E^0$ is given by
\begin{eqnarray}
S_K^0 & = & \int_0^\infty\ r^3 dr \, \frac{1}{2} \dot\phi_b^2 \\
S_V^0 & = & \int_0^\infty\ r^3 dr \, V_{e^4}(\phi_b) 
\end{eqnarray}
where $\phi_b$ minimizes $S_E^0$. Now rescale $ x^\mu = \lambda r^\mu$, so that 
\begin{eqnarray}
S_K^0 & = & \frac{1}{\lambda^2} \,I_K  \\
S_V^0 & = & \frac{1}{\lambda^4}\, I_V\ \ \ ,
\end{eqnarray}
where
\begin{eqnarray}
I_K & = &  \int_0^\infty\ x^3 dx \, \frac{1}{2} \dot\phi_b^2 \\
I_V & = &  \int_0^\infty\ x^3 dx \, V_{e^4}(\phi_b) \ \ \ .
\end{eqnarray} 
Thus,
\begin{equation}
S_E^0 = \frac{1}{\lambda^2} \,I_K + \frac{1}{\lambda^4}\, I_V \ \ \ .
\end{equation}
Now, 
\begin{equation}
-\frac{d S_E^0}{d\lambda} = \frac{2}{\lambda^3} \,I_K + \frac{4}{\lambda^5}\, I_V \ \ \ .
\end{equation}
Since $\phi_b$ extremizes $S_E^0$ we must have that 
\begin{equation}
\label{eq:extrema}
\left(\frac{d S_E^0}{d\lambda}\right)_{\lambda=1} = 0 \ \ \ .
\end{equation}
Thus, we obtain $I_K=-2 I_V$, so that $S_K^0=2 S_E^0=-2S_V^0$.

It is now straightforward to see how the scaling relations can be broken by the inclusion of higher order contributions to the effective action. First, recall that Eq.~(\ref{eq:extrema}) only holds for an extremal field configuration. In the present context, the bounce solution $\phi_b$ extremizes $S_E$ in the presence of $V_{e^4}$ but not when next-to-next-to leading order contributions are included via $Z_{e^2}$ and $V_{e^6}$. Second -- and more generally -- should the effective action receive contributions beyond second order in derivatives, $S_K$ will no longer scale as $1/\lambda^2$ upon the change of variables $x^\mu=\lambda r^\mu$. 
}

\bibliographystyle{JHEP}
\bibliography{mybib}

\end{document}